# Individual-specific precision neuroimaging of learning-related plasticity

**Authors:**

Simon Leipold [1,2] & Ryssa Moffat [1]

**Affiliations:**

[1] Social Brain Sciences Lab, Department of Humanities, Social and Political Sciences, ETH Zurich

[2] Neuroscience Center Zurich, University of Zurich and ETH Zurich

**Corresponding author:**

Simon Leipold
ETH Zürich
Social Brain Sciences Lab
Stampfenbachstrasse 69
8092 Zürich
Switzerland

**Email:**

simon.leipold@gess.ethz.ch

**Keywords:**

precision functional mapping, longitudinal, fMRI, fNIRS, within-subject analysis, deep phenotyping, individual differences

## Abstract

Studying learning-related plasticity is central to understanding the acquisition of complex skills, for example learning to master a musical instrument. Over the past three decades, conventional group-based functional magnetic resonance imaging (fMRI) studies have advanced our understanding of how humans' neural representations change during skill acquisition. However, group-based fMRI studies average across heterogeneous learners and often rely on coarse pre- versus post-training comparisons, limiting the spatial and temporal precision with which neural changes can be estimated. Here, we outline an individual-specific precision approach that tracks neural changes within individuals by collecting high-quality neuroimaging data frequently over the course of training, mapping brain function in each person's own anatomical space, and gathering detailed behavioral measures of learning, allowing neural trajectories to be directly linked to individual learning progress. Complementing fMRI with mobile neuroimaging methods, such as functional near-infrared spectroscopy (fNIRS), will enable researchers to track plasticity during naturalistic practice and across extended time scales. This multi-modal approach will enhance sensitivity to individual learning trajectories and will offer more nuanced insights into how neural representations change with training. We also discuss how findings can be generalized beyond individuals, including through statistical methods based on replication in additional individuals. Together, this approach allows researchers to design highly informative longitudinal training studies that advance a personalized account of skill learning in the human brain.

## 1. Introduction

The brain adapts and changes in response to experience across the lifespan, a capacity known as neural plasticity. Studying neural plasticity is essential for understanding how humans acquire complex skills, such as playing a musical instrument or learning a foreign language (Lövdén et al., 2020). Neural plasticity has been a central focus in neuroscience ever since the discovery that the brain remains malleable in adult animals, including humans (Kaas, 1991).

Neural plasticity can be studied at multiple levels, from synapses to large-scale brain networks (Akbaritabar and Rubin, 2024), and occurs during learning, development, and in response to brain damage (Hille et al., 2024; Pascual-Leone et al., 2005). Here, we focus on learning-related neural plasticity in humans, which we define as changes in **neural representations** that correlate with learning (Buonomano and Merzenich, 1998). More specifically, we focus on learning of complex **skills**, defined as abilities that improve with practice (Rosenbaum et al., 2001). Complex skills encompass both perceptual-motor skills, such as playing the piano, and cognitive skills, such as playing chess. Changes in neural representations are functional in nature, but are thought to reflect underlying structural changes associated with synaptic plasticity (Lövdén et al., 2010; Paillard, 1976; Will et al., 2008). These microscopic changes cannot be directly resolved with current non-invasive functional neuroimaging in humans (Hille et al., 2024). Nonetheless, structural plasticity can be observed at a larger scale as changes in gray and white matter measured with structural and diffusion-weighted neuroimaging (Bezzola et al., 2011; Draganski and May, 2008; Wenger et al., 2017a). Several aspects of the framework outlined below may also inform research using structural imaging modalities, although we do not discuss these modalities in detail.

Understanding how the acquisition of complex, culturally embedded skills such as music or dance reshapes neural representations is best achieved by studying plasticity directly in the *human* brain, rather than relying solely on non-human animal models (Cross, 2025; Münte et al., 2002; Zatorre, 2013). Over the past three decades, researchers have investigated learning-related plasticity in humans using functional neuroimaging techniques, most prominently functional magnetic resonance imaging (fMRI). Although this approach presents several methodological challenges (Poldrack, 2000), fMRI remains the only non-invasive method for tracking plasticity across the whole human brain at millimeter resolution.

Longitudinal fMRI studies have provided important insights into how neural representations change over time as individuals acquire new skills. In a conventional group-based fMRI study of learning, a group of participants is scanned before and after acquiring a perceptual, motor, or cognitive skill (**Figure 1A**). The fMRI data is **spatially normalized** to a common template, and statistical analyses aim to identify group-level changes in **blood-oxygenation-level-dependent (BOLD) signal** from pre- to post-training. Learning-related changes in BOLD signal have been studied across a range of domains (Chein and Schneider, 2005; Hardwick et al., 2013; Kelly and Garavan, 2005), including perceptual skill learning in the visual (Gauthier et al., 1999; Poldrack, 1998) and auditory modalities (Jäncke et al., 2001; Zatorre et al., 2012), observational and physical motor skill learning (Apšvalka et al., 2018; Bassett et al., 2015, 2011; Berlot et al., 2020; Cross et al., 2009; Gardner et al., 2017; Karni et al., 1995), and cognitive skill learning (Park et al., 2024; Stewart et al., 2003). These group-based designs have yielded important insights into how neural activation changes with learning. In motor skill learning, for example, group-based studies have consistently implicated a distributed network involving premotor and supplementary motor areas, parietal cortex, and subcortical and cerebellar structures (Hardwick et al., 2013). Yet the direction and temporal profile of learning-related BOLD changes within this network remain less clear, with studies reporting both increases and decreases over the course of training (Berlot et al., 2020; Steele and Penhune, 2010; Wiestler and Diedrichsen, 2013). These inconsistencies may partly reflect the difficulty of distinguishing different stages of learning within an individual from genuinely different trajectories across individuals, as well as from artifacts introduced by group averaging.

In this paper, we outline the limitations of conventional group-based neuroimaging studies of learning-related plasticity and discuss how individual-specific precision approaches can address these challenges. We propose conceptual and methodological advances that build on established principles of **precision neuroimaging** (Gordon et al., 2017; Gratton and Braga, 2025; Michon et al., 2022) to strengthen our understanding of learning-related plasticity in individuals. We also identify key challenges for applying precision neuroimaging to human skill learning and conclude by comparing group-based and precision approaches in practice. Our aim is not to argue that precision approaches are universally preferable, but rather that they are particularly well suited to research questions centered on within-person learning trajectories and their neural foundations.

The methodological advantages of individual-specific precision approaches apply broadly to the investigation of within-individual brain-behavior relationships across domains (Gratton et al., 2022; Gratton and Braga, 2025; Vinci-Booher et al., 2025). These advantages can also be leveraged to study learning-related plasticity, where changes in behavior and neural activation must be tracked in tandem across extended periods of time.

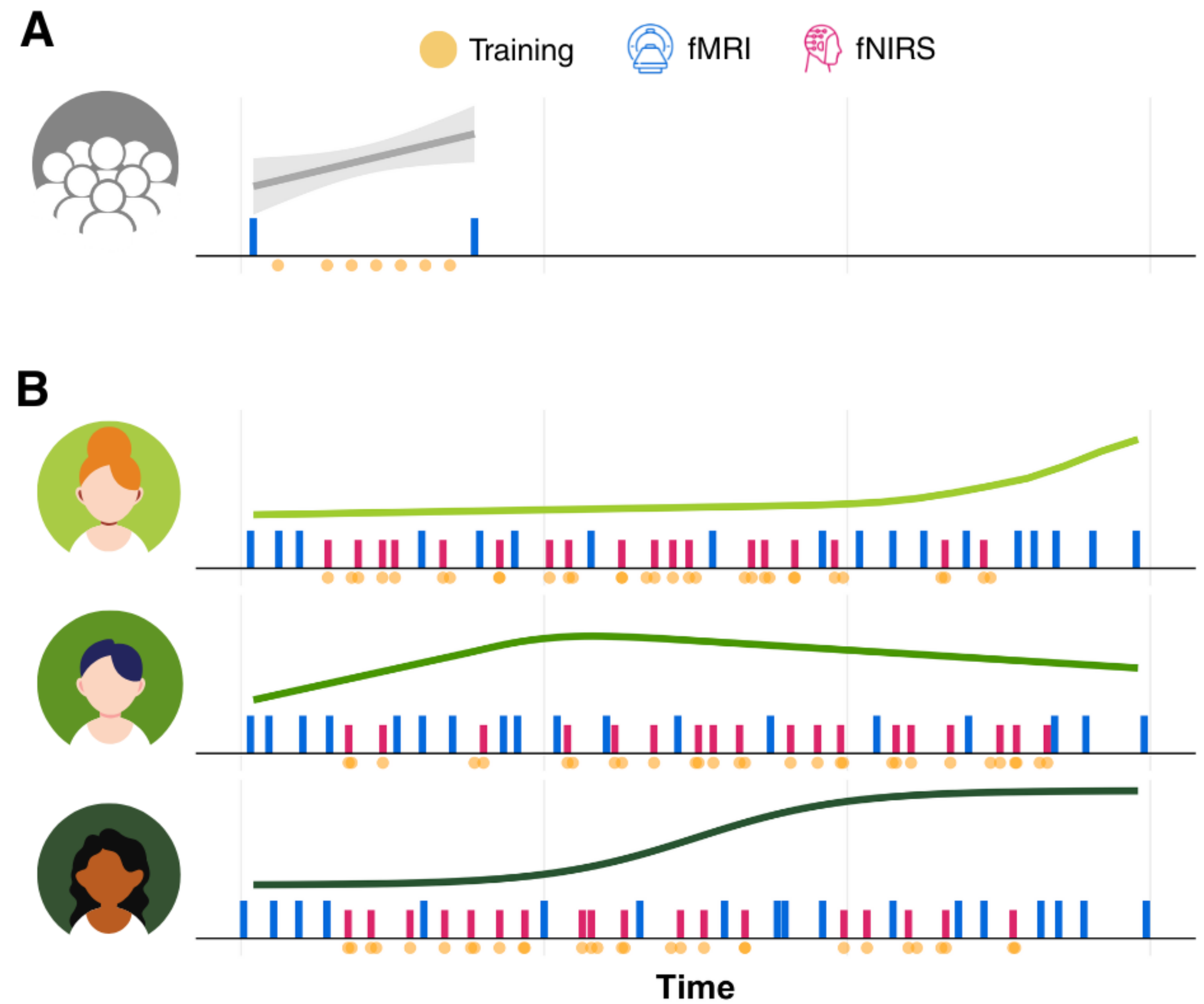

**Figure 1. Group-based versus individual-specific precision designs for studying learning-related neural plasticity.**

(A) Conventional group-based studies typically assess learning-related change across a few measurement points (e.g., before and after training). Averaging across individuals yields an overall learning trajectory (grey line) but obscures individual variability in learning dynamics. (B) Individual-specific, densely sampled designs capture within-person trajectories across extended periods. Repeated behavioral training sessions (yellow dots) are complemented by frequent neuroimaging assessments using functional magnetic resonance imaging (fMRI; blue bars) and potentially mobile modalities such as functional near-infrared spectroscopy (fNIRS; pink bars). This approach allows the characterization of person-specific learning curves (green individual lines) and their temporal coupling with neural change, providing a richer account of learning-related plasticity than conventional group averages.

## 2. Why should we study learning-related plasticity at the level of the individual?

Group-based approaches to studying learning-related plasticity average across participants to derive group-level effects, which can obscure individual differences. These differences include variation in (functional) neuroanatomy, task strategies, and learning trajectories. Averaging across heterogeneous individuals can yield group-level effects that are not representative of most participants. Moreover, relationships between neural and behavioral changes observed at the group level may not align with those within individuals, as interindividual heterogeneity in neural activity and behavior can reshape these relationships. Ultimately, these issues reflect a lack of group-to-individual generalizability, known as **non-ergodicity**, that is common in the fields of cognitive neuroscience and psychology (Fisher et al., 2018; Hunter et al., 2024; Mattoni et al., 2025; Molenaar, 2004; Molenaar and Campbell, 2009). In this context, the cognitive and behavioral sciences have seen growing interest in individual-focused, small-*N* approaches that prioritize understanding change within individuals, rather than drawing conclusions primarily from group averages (Normand, 2016; Smith and Little, 2018). Individual-specific neuroimaging approaches build directly on this logic by tracking changes within each participant, providing a principled way to capture learning-related plasticity at the level at which plasticity occurs while maintaining sensitivity to individual differences (Gratton and Braga, 2025).

### 2.1. Preserving spatial precision through analysis in native anatomical space

A central challenge in group-based neuroimaging studies is the substantial anatomical variability across individuals. Brains differ between individuals at all levels, from cytoarchitectonic organization (Amunts et al., 2020) to cortical folding (Mangin et al., 2004) and white-matter tract organization (Bürgel et al., 2006). This inter-individual variability is so pronounced that individuals can be accurately identified from their anatomy (Valizadeh et al., 2018) or connectivity patterns (Yeh et al., 2016). Group-based studies typically address such inter-individual differences with spatial normalization to a standard template of the human brain, but these normalization algorithms remain imperfect (Klein et al., 2009). Residual misalignment is often mitigated by spatial smoothing, yet both smoothing and the interpolation inherent to normalization reduce spatial specificity (Stelzer et al., 2014). This is particularly problematic for studies of plasticity, where detecting consistent learning-related changes requires precise spatial alignment between participants.

Individual-specific approaches address this problem by analyzing each participant's brain in **native space**, preserving fine-grained anatomical detail and avoiding spatial normalization. This allows for anatomically precise tracking of learning-related changes in neural representations and increases sensitivity to localized plasticity that may vary across individuals.

### 2.2. Capturing individual functional anatomy with localizer tasks

An even greater challenge than anatomical variability in group-based neuroimaging is inter-individual variability in *functional* anatomy, meaning that the same functional area may not occupy the same anatomical location across individuals (Fedorenko, 2021; Frost and Goebel, 2012). This undermines the assumption that spatial normalization can align functionally corresponding regions across brains. While primary sensory and motor areas show relatively consistent mapping, higher-level cognitive functions, particularly in frontal and parietal cortices, exhibit much weaker structure-function correspondence across individuals (Fedorenko et al., 2012, 2010; Tahmasebi et al., 2012).

Individual-specific analyses address this issue by defining regions of interest based on functional responses in each brain. Because these analyses are often conducted in native space, standard coordinates cannot be used to infer function (Brett et al., 2002). Instead, individualized functional localizers provide a powerful alternative: functional localizers are experimental tasks designed to identify brain regions involved in specific perceptual, motor, or cognitive functions (Saxe et al., 2006). Localizers define regions by contrasting conditions that engage the target function with those that do not. They are well established in visual neuroscience, for example in localizing face-responsive regions by contrasting faces with objects (Kanwisher et al., 1997), and in language research, where they identify regions robustly activated by linguistic input (Fedorenko et al., 2010). Identifying functionally defined regions within individuals increases sensitivity to learning-related plasticity by enabling researchers to track changes in functionally equivalent areas over time. For example, a simple finger-tapping task can be used to localize hand areas in each participant's primary motor cortex. Once defined, these individual-specific regions become the focal point of repeated measurements of brain activity to track how motor skill learning across training sessions reshapes functional responses in the identified regions.

### 2.3. Accounting for cognitive strategy differences across individuals

The same cognitive function can be supported by different brain regions or networks, a principle known as **degeneracy** (Edelman and Gally, 2001; Price and Friston, 2002). A theory explaining degeneracy is that, for a given task used to test a specific cognitive function, individuals can rely on different task strategies, thereby activating different brain regions or networks (Seghier and Price, 2018). Across domains such as language and memory, comparisons of group-average brain maps and individual-specific brain maps reveal large mismatches in the location of peak activation, beyond what would be expected from functional-anatomical variability or spatial normalization errors (Heun et al., 2000; Miller et al., 2002; Sanfratello et al., 2014). These discrepancies can be interpreted as reflecting strategy differences (Nadeau et al., 1998), a view supported by evidence that inter-individual variability in activation patterns is stable over time and therefore not attributable to measurement noise (Miller et al., 2012, 2002; Sanfratello et al., 2014). Similar variability in task strategies as in humans has also been observed in rodents (Gallero-Salas et al., 2021; Gilad et al., 2018; Piet et al., 2024). Group-based analyses risk averaging across participants, who use different strategies to complete tasks, thereby obscuring meaningful individual activation patterns that result from specific strategies.

This issue is especially important in studies of neural plasticity, where the learning process itself may involve changes in task strategy. In a motor skill learning study, for example, some participants shifted from implicit to explicit strategies, each supported by distinct neural systems (Grafton et al., 1995). Similar shifts occur in cognitive learning, where individuals may spontaneously switch strategies (Schuck et al., 2015), such as moving from spatial to nonspatial strategies during navigation (Iaria et al., 2003). If strategy changes occur during training in some individuals but not others, group-level analyses average across incompatible neural activation patterns, and the point in time at which they became incompatible is not accounted for in the analysis. In other words, when individuals adopt different strategies at different stages of training, the group-average trajectory becomes a blend of distinct learning patterns. This blending can obscure meaningful activation changes, cancel out genuine effects, or produce misleading maps of brain activity that do not reflect any individual trajectory.

These limitations highlight the need to track learning trajectories and strategies in individuals to interpret learning-related plasticity accurately. By focusing on within-subject trajectories rather than group averages, individual-specific approaches reveal how people differ in the ways they approach and adapt to tasks. Strategy use can be assessed through behavioral markers (Gilad et al., 2018), verbal reports (Grafton et al., 1995), or model-based analyses of behavior (Yang and Jiang, 2025), and then directly related to brain activity for each individual. This enables researchers to determine whether neural activation changes reflect plasticity alongside a stable strategy or recruitment of different neural systems as strategy shifts occur.

### 2.4. Revealing individual learning trajectories over time using individual-specific models

Learning is shaped by shared principles, yet differs across individuals in its pace and form (Anderson et al., 2021; Bonte and Brem, 2024; Stern, 2017). People vary in their initial levels of knowledge and skill, the rate at which they progress (Zerr et al., 2018), the shape of their learning curves, and the level of proficiency they ultimately achieve. These behavioral differences are mirrored in learning-related neural plasticity, which often follows nonlinear trajectories with distinct phases (Karni et al., 1998, 1995). As a result, individuals completing the same training may pass through learning phases in different orders and rates, where each phase is associated with distinct brain activation patterns (Sakai et al., 1998). Averaging across such variability blends different neural states, obscuring the diversity of neural processes that might underlie individuals' behavioral learning trajectories. Individual-specific approaches preserve these differences, enabling distinct behavioral trajectories to be linked to the corresponding neural changes. This is particularly important because individual differences in learning trajectories are not random: For example, baseline abilities shape both the magnitude and nature of change with training, a phenomenon known as **aptitude-treatment interactions** (Anderson et al., 2021; Cronbach, 1975). Individual-specific trajectories can reveal how initial abilities shape behavioral improvements, as well as the neural plasticity that supports them.

A central aim of neural plasticity research is to model neural trajectories as a function of non-neural variables that capture learning or training progress, most often behavioral measures such as accuracy, response times, eye tracking, or kinematics. These brain-behavior relationships can be modeled at the group level to identify population trajectories but may fail to characterize any one individual's trajectory (Hunter et al., 2024). For example, a group-level analysis might show that accuracy increases while brain activation decreases across training, suggesting a negative brain-behavior relationship. Yet for some individuals, increases in accuracy may be accompanied by increases in activation early in training and decreases in activation at later stages. The averaged group-level trend obscures how distinct within-subject trajectories unfold over time. Individual-specific modeling is therefore essential to capture the individual trajectories that link behavior to neural plasticity.

## 3. How can we measure learning-related plasticity with high precision?

Longitudinal group-based neuroimaging studies of learning-related plasticity are inherently time and resource intensive. To achieve sufficient statistical power, 20 to 30 or more participants' brains are typically scanned at least twice, at least once before and once after training, with the intervening training often spanning multiple days to weeks to months. To make such studies feasible, researchers often must compromise on scan duration or the number of measurements – compromises which reduce **measurement precision**. Measurement precision is essential for high **test-retest reliability** (Nebe et al., 2023), and reliable measurements are required to distinguish changes in brain activation induced by training from spurious signal fluctuations caused by noise. Individual-specific approaches address this challenge by studying fewer participants' learning trajectories in greater depth and reallocating resources toward repeated measurements within individuals. This represents a different allocation of resources rather than a reduction in overall resource demands.

### 3.1. Increasing test-retest reliability through extended sampling and denoising

With fMRI, learning-related plasticity can be said to have occurred when systematic changes in neural representations track with learning progression (Buonomano and Merzenich, 1998). However, fMRI signals such as the BOLD response are inherently noisy, and fluctuations introduced by noise can obscure true learning effects. Sources of noise include scanner-related factors such as thermal noise and instability, head motion, physiological fluctuations such as heartbeat and respiration, variations in subject state such as wakefulness or caffeine intake, and inconsistencies in scanner setup or head placement across sessions (Greve et al., 2013; Liu, 2016). These factors reduce measurement precision (Nebe et al., 2023), lowering the test-retest reliability of the measures across sessions and limiting the likelihood of detecting subtle learning-related changes in neural representations over time. Consequently, many commonly used fMRI measures, including task activation and functional connectivity, show poor test-retest reliability at the trial numbers and task durations typical of conventional group-based studies (Elliott et al., 2020; Noble et al., 2019). For example, a large-scale meta-analysis found that widely used task-based activation measures, such as amygdala activation during emotion processing or dorsolateral prefrontal cortex activation during an n-back working memory task, fall well below accepted reliability standards (Elliott et al., 2020). As a result, apparent changes in brain activation may reflect noise rather than true neural plasticity, a limitation that is particularly problematic for longitudinal training designs aimed at detecting subtle learning-related changes in brain function.

One way to increase test-retest reliability in fMRI studies is to collect more data from each individual, often called **extended sampling** or dense sampling (Gordon et al., 2017; Gratton et al., 2022; Laumann et al., 2015). In task-based designs, this means increasing the number of trials per condition, often by a factor of five to ten relative to conventional group-based studies (Elliott et al., 2021). Averaging over more trials reduces noise and stabilizes the estimates of brain activation, which is crucial for longitudinal studies of learning-related plasticity where the aim is to detect true learning-related changes in a given individual over extended timescales. For resting-state functional connectivity, about 40 minutes of high-quality data is needed to achieve stable estimates, four to ten times longer than typical group-based studies acquire (Gordon et al., 2017). Some functional connectivity-derived metrics are inherently more reliable than others, meaning that the optimal duration depends on the selected metric (Gordon et al., 2017). Because extended data collection in a single session can be burdensome for participants, brain scanning is usually distributed across multiple sessions. In the context of learning studies, this implies that robust baseline estimates require repeated measurement sessions to adequately quantify baseline reliability (**Figure 1B**).

A second way to increase the reliability of fMRI measures is through **denoising**. Denoising involves removing artifacts and random fluctuations from the signal. Beyond conventional acquisition and preprocessing steps for reducing noise (Liu, 2016; Nebe et al., 2023; Power et al., 2014), data quality can be further improved with more advanced methods: These include **multi-echo acquisition** combined with denoising (Kundu et al., 2017), which lowers the amount of data needed for reliable estimates (Lynch et al., 2020), and real-time motion tracking, which allow the number of runs or resting-state volumes to be tailored to session-specific data quality, increasing the proportion of usable data while minimizing unnecessary scan time (Dosenbach et al., 2017). By combining extended sampling with advanced denoising strategies, fMRI measurements achieve higher test-retest reliability and greater sensitivity to true learning-related changes. Because test-retest reliability is essential for robust findings, reliability should be routinely quantified and reported for each dependent variable in longitudinal fMRI studies of learning (Elliott et al., 2020; Moriarity and Alloy, 2021; Parsons et al., 2019).

### 3.2. Tracking learning dynamics using multi-session longitudinal data

Group-based longitudinal fMRI studies of learning-related plasticity in humans are inherently time- and resource-intensive, given the high cost and logistical demands of repeated brain scanning and the need to implement training regimes that often extend over days or longer. As a result, researchers must balance the number of participants with the depth of sampling within each participant (Naselaris et al., 2021; Ooi et al., 2025). As described above, many studies adopt a pre-training versus post-training design, which efficiently captures the brain states underlying untrained and trained behaviors but leaves the trajectory of plastic change in the brain unobserved. This is a critical limitation of studies with only two timepoints, as learning and its neural substrates evolve dynamically and often nonlinearly (Wenger et al., 2017b). Neural and behavioral changes associated with learning typically unfold in distinct phases rather than along a smooth continuum (Karni et al., 1998, 1995). In motor skill learning, for instance, rapid early improvements are followed by slower learning phases, each supported by partly dissociable neural mechanisms (Doyon and Benali, 2005; Penhune and Steele, 2012). Thus, designs limited to pre- and post-testing therefore risk obscuring crucial intermediate stages of learning-related plasticity. Recent studies have started to address this limitation by increasing the number of sessions (e.g., Berlot et al., 2020; Garzón et al., 2023; Matuszewski et al., 2021). However, precisely characterizing the temporal dynamics of learning-related plasticity might require an even greater number of measurement points throughout the training period and after training has ceased (**Figure 1B**).

To overcome the limitations of group-based designs, which often do not sample the intermediate stages of plastic change, proponents of precision neuroimaging emphasize dense multi-session data collection in a small number of individuals (Vinci-Booher et al., 2025). By allocating resources to repeated measurements from few participants rather than fewer measurements from large numbers of individual participants (Normand, 2016), the precision neuroimaging approach affords detailed longitudinal characterization of baseline brain function, learning trajectories, and post-training outcomes. A recent study applied this method to investigate working memory representations in the lateral prefrontal cortex of three individuals who practiced a working memory task and a serial reaction time task over three months, with more than 20 fMRI sessions per participant (Miller et al., 2022). The design enabled the authors to reconcile conflicting findings between human and nonhuman primate research, where nonhuman primates are typically extensively trained before neural recordings begin. Similarly, Newbold et al. (2020) collected as many as 63 resting-state scans from each of three individuals during a period of arm immobilization. The intervention induced large-scale reorganization of sensorimotor brain networks, with decreased functional connectivity linked to the immobilized limb. Both of these studies demonstrate how dense, individual-specific sampling can uncover otherwise inaccessible insights into experience-dependent plasticity in the human brain, an approach that holds great promise for capturing the fine-grained dynamics of learning-related plasticity.

### 3.3. Linking behavioral and neural learning trajectories through deep phenotyping

The availability of densely sampled neural learning trajectories in a small number of individuals requires equally dense sampling of behavior to establish meaningful links between behavioral and neural change. Just as the test-retest reliability of neural data improves with repeated measurements, the reliability of behavioral measures derived from cognitive tasks increases when aggregated across a large number of trials (Lee et al., 2025). The strategy of collecting large and diverse within-individual datasets on the brain, behavior, and other domains has been termed deep sampling or deep phenotyping (Kupers et al., 2024; Naselaris et al., 2021; Poldrack et al., 2015). A landmark example is a study that followed a single individual over 18 months while collecting extensive neuroimaging, behavioral, self-report, physical health, and genetic data (Poldrack et al., 2015). Precision studies of learning-related plasticity can capture rich longitudinal datasets on learning behavior and complementary non-neural measures, generating dense time series that can be directly related to neural plasticity trajectories within each individual (Triana et al., 2024). Relevant non-neural measures may include, but are not limited to sleep, which plays a critical role in memory consolidation through both its quantity and quality (Brodt et al., 2023), and hormone levels (Goldstein-Piekarski et al., 2020; Pritschet et al., 2020), which can alter memory and learning outcomes (Taxier et al., 2020). In addition, assessments of psychological states that modulate learning, such as mood, motivation, or boredom, may provide valuable complementary information.

Beyond descriptive links between behavior and neural change, densely sampled individual-level datasets are particularly well suited for computational modeling approaches (Dupré la Tour et al., 2025; Forstmann et al., 2016; Naselaris et al., 2021; Wichmann and Hill, 2001; Wu et al., 2006). Repeated behavioral measurements enable fitting models such as psychometric functions or evidence accumulation models at the level of individual participants, allowing underlying computational parameters to be tracked as they evolve with experience and practice (Cochrane et al., 2023; Parker et al., 2025). Framing learning-related plasticity in terms of changes in computational states provides a principled way to link behavioral adaptation to neural dynamics and opens the door to joint modeling of behavioral, neural, and physiological measures (Love, 2024; Mathis and Mathis, 2025). Such approaches offer a promising avenue for future work aimed at understanding learning as plasticity in computational processes, rather than solely as changes in observed performance.

### 3.4. Extending plasticity research to everyday-life learning contexts

In an ideal scenario, it would be possible to investigate neural changes as learning occurs, capturing the dynamics of plasticity during the learning process itself. However, this is rarely feasible with fMRI, as the MRI scanner environment strongly constrains movement and naturalistic behavior. These constraints are particularly problematic when studying the acquisition of complex motor skills, such as learning to dance the tango or playing the drums, which inherently involve extensive motion and long training periods. Increasing the number of fMRI sessions cannot fully overcome this issue, since most learning occurs outside the MRI scanner.

A promising way forward lies in the use of mobile neuroimaging technologies such as electroencephalography (EEG) and functional near-infrared spectroscopy (fNIRS) (**Figure 1B**). Mobile neuroimaging allows researchers to track learning-related plasticity during more naturalistic behavior during training sessions. The combination of fMRI and fNIRS is particularly advantageous, as both techniques measure hemodynamic signals related to blood oxygenation with similar temporal resolution, and neural activity is modelled using the hemodynamic response function (Strangman et al., 2002). Anatomical information acquired using fMRI functional localizers can guide fNIRS sensor placement during learning tasks outside the MRI scanner to optimize coverage of individual-specific regions of interest and signal quality. Moreover, adopting a precision approach with a small number of intensively studied participants allows for practical and methodological refinements. For example, individual anatomical MRI scans can serve as personalized head models for fNIRS signal modeling. Custom 3D-printed sensor caps can be tailored to each participant (Lühmann et al., 2024) and paired with computer vision approaches to position the cap precisely each session (Rahimpour Jounghani et al., 2025), thereby ensuring high inter-session consistency. When applied longitudinally, fNIRS enables the tracking of neural changes across extended periods of naturalistic practice, thereby offering the opportunity to link laboratory-based measures of plasticity to everyday learning contexts. Its applicability to **hyperscanning**, measuring brain signals from two (or more) individuals at the same time, further opens opportunities to examine social and interactive learning, where multiple individuals learn together or from each other (Moffat et al., 2025, 2024).

## 4. Challenges in applying precision neuroimaging to learning

Implementing an individual-specific precision approach to study learning-related plasticity poses several challenges. A key difficulty lies in disentangling genuine short-term learning-related change from apparent instability in neural and behavioral measures, which can be misinterpreted as low test-retest reliability. Importantly, this challenge does not reflect a lack of measurement precision per se, but rather the systematic changes introduced by learning itself. A second consideration, inherent to approaches that prioritize individuals over groups (Gordon et al., 2017; Naselaris et al., 2021), concerns generalization of findings beyond the specific participants studied. This is only a limitation insofar as population-level inference is the primary goal. When individual-specific designs instead prioritize accurate within-person estimation, reduced population generalizability reflects a deliberate trade-off rather than a shortcoming.

### 4.1. Quantifying reliable change in the presence of learning

A central goal of studying learning-related neural plasticity is to capture genuine change over time. Yet this focus on *change* creates a methodological paradox for the precision approaches, which are centered around high measurement reliability, the *consistency* of measures across repeated assessments. The paradox can be summarized as follows: as individuals learn, their neural and behavioral responses are expected to change, but these true changes may be comparable in magnitude, or even obscured by, low test-retest reliability. Thus, high test-retest reliability is essential for detecting genuine change. Yet short-term learning introduces within-subject variability that undermines the apparent stability of behavioral and neural measures. In other words, test-retest reliability assumes that the same condition is measured repeatedly, whereas learning continuously alters the very conditions under which those measures are obtained.

This issue is particularly pronounced for skills that can be mastered within a single fMRI scanning session, such as the learning of simple motor sequences (Karni et al., 1995). For such rapid forms of learning, techniques that capture neural change on short timescales, such as EEG or trial-evoked task-based fMRI responses, may be better suited than measures such as functional connectivity that often require longer sampling windows. Rapid learning-related changes may unfold across short task epochs and may therefore be less well captured by measures that depend on averaging over extended periods.

Individual-specific precision approaches are best suited for investigating forms of learning that unfold gradually over extended periods or that involve complex and difficult tasks, maximizing the length of the window within which dense multi-session data collection can occur. Put differently, the frequency of measurements across a period of learning is a sampling rate that determines the temporal resolution of the experimental design. The sampling rate must be high enough to capture fine-grained changes in the learning time series. This ultimately means that sampling rates of in-lab sessions are more likely to be feasible and to be appropriately scaled for learning trajectories that develop over days and weeks rather than seconds or minutes. For example, mastering a musical instrument requires sustained practice and may follow a trajectory of continuous improvement over months or years (Leipold et al., 2021), often punctuated by plateaus and sudden gains in performance. Such long-term, nonlinear learning trajectories are ideally suited to an individual-specific precision approach, which can capture the detailed temporal dynamics of change within each learner.

### 4.2. Balancing individualization and generalization

An individual-specific precision approach for studying learning involves making statistical inferences at the level of each participant rather than at the group level. Each participant's data is analyzed independently, and replication occurs across individuals within the same study: the data from one participant serve as an independent replication of another's findings (Schwarzkopf and Huang, 2024). In the individual-specific approach, multiple participants provide multiple independent replication opportunities: If a study includes five participants and the same effect is observed in all five, this constitutes a fivefold replication at the individual level (Smith and Little, 2018). To support transparent interpretation, statistical results should be accompanied by visualizations of each participant's data, highlighting both shared and idiosyncratic patterns. Transparency can be further strengthened through preregistration of hypothesized results (Nosek et al., 2018). For example, after identifying learning-related effects in an initial participant, simulations based on that dataset can guide preregistration of expected effects and sampling decisions for subsequent participants before collecting and analyzing their data.

Establishing reliable effects within a single individual requires repeated observations of the phenomenon of interest across many trials (Normand, 2016). These within-person repetitions provide the sampling basis for inferential statistics. This logic, common in psychophysics and non-human animal electrophysiology, contrasts with the conventional group-based framework in which statistical inference is drawn about a population from a sample of individuals. The within-individual approach, however, does not permit formal generalization to a statistical population as is customary in group-level analyses that use frequentist null-hypothesis significance testing to infer population effects from sample data (Smith and Little, 2018). Yet formal generalization may not always be required, particularly when the scientific aim is to understand mechanisms that operate within individuals (Mattoni et al., 2025; Naselaris et al., 2021). In such cases, the goal is not to estimate the average effect across a population, but to characterize the principles and dynamics that govern individual trajectories of learning or plasticity. Insights derived from detailed, within-person analyses can reveal how learning unfolds in real time, how neural and behavioral adaptations co-evolve, and how these processes differ across individuals. Importantly, comparing individual trajectories can also uncover recurring qualitative patterns that provide a principled way to summarize similarity and distinctiveness across individuals. These patterns can, in turn, generate new hypotheses and inform theories of learning that generalize conceptually across people.

At the same time, emerging statistical frameworks such as **population prevalence inference** offer a way to bridge individual-specific precision with formal generalization (Ince et al., 2022). Instead of testing for an average effect at the population level, prevalence inference estimates the proportion of individuals in the population who exhibit a given effect, based on its occurrence across the studied individuals. Crucially, this approach requires studies to be sufficiently powered to detect effects within individuals by collecting an adequate number of observations per participant. Both frequentist (Allefeld et al., 2016; Rosenblatt et al., 2014) and Bayesian (Ince et al., 2021) formulations of population prevalence inference have recently been developed and applied in cognitive neuroscience. The prevalence approach retains the strength of individual-level analysis while offering a principled route toward population-level inference grounded in empirical replication across individuals.

## 5. Implementing longitudinal precision neuroimaging in practice

To illustrate how conventional group-based designs and individual-specific precision approaches differ in practice, we first present a hypothetical longitudinal fMRI study of learning-related plasticity as a worked example. We then discuss hybrid designs that combine elements of both approaches and may help balance their respective strengths. **Table 1** summarizes the main differences between group-based and precision designs.

### 5.1. A practical comparison of group-based and precision approaches

To contrast group-based and precision approaches, we consider a study of auditory category learning adapted from Jiang et al. (2018). The study investigates how neural representations change as individuals learn to classify unfamiliar sounds into novel categories. In auditory perception, stable categories must be formed despite substantial acoustic variability: sounds from the same category can differ considerably, whereas sounds from different categories can be highly similar. In the following, we use this example to illustrate how group-based and precision approaches differ in study design, research questions, and interpretation.

#### *5.1.1. Participants*

In a group-based design, the study recruits a larger sample of participants from the population to which the findings are intended to generalize (e.g., 20 to 30 participants). In contrast, a precision design includes only a small number of participants (e.g., three to five), because each individual is studied intensively across repeated sessions. Participant selection therefore needs to consider both the required time commitment and the likelihood of completing dense longitudinal sampling with high adherence.

#### 5.1.2. Stimuli

In our hypothetical study, the group-based design starts from two prototype animal calls, each representing a different category, to generate a set of novel auditory stimuli. Animal calls are useful for studying auditory category learning because they are natural vocalizations yet sufficiently unfamiliar to human listeners that learning can be examined under controlled conditions. Intermediate sounds are then created that vary gradually between the two categories, for example using a morphing algorithm. From this larger set of intermediate sounds, individual stimuli are selected and presented in pairs during fMRI. The pairs are chosen so that some contain sounds that are equally different acoustically, while either belonging to the same category or to different categories. This makes it possible to test whether the learning process changes representations of acoustic features, of sound categories, or both. In a precision design, the same logic applies, but the stimulus set is expanded (Kupers et al., 2024; Naselaris et al., 2021). Extended sampling makes it feasible to include more exemplars per category, sample a wider range of intermediate sounds, and examine more categories. This reduces the influence of stimulus-specific features and allows for tests of generalization to new sounds.

#### 5.1.3. fMRI tasks

For both designs, participants perform an attentionally demanding auditory task during fMRI in which they respond to a stimulus feature unrelated to category membership. This keeps them engaged while helping to isolate neural representations of the sounds from activation related to explicit categorization and decision demands. In a precision design, the task is used with more trials and repeated stimulus presentations to improve the reliability of individual-level activation patterns. Because more scanning time is available per participant, a parallel visual categorization task might be included to test whether similar plasticity principles extend across sensory modalities (Jiang et al., 2007).

#### 5.1.4. Study design

The group-based study involves a pre-training fMRI session, several hours of behavioral category training completed at home, and a post-training fMRI session **(Figure 1A)**. Training uses a feedback-based categorization task that begins with easy stimuli and gradually introduces more difficult stimuli closer to the category boundary as performance improves. In a precision study, participants complete repeated baseline scans before training to quantify the reliability of the neural activations associated with the sound stimuli, and training is interleaved with repeated fMRI sessions across learning **(Figure 1B)**. This allows neural change to be tracked at multiple stages of category acquisition. Mobile methods such as fNIRS or EEG can be used to capture learning-related processes while participants practice the categorization task.

#### 5.1.5. Research Questions

In a group-based design, the main question is whether category training changes neural representations from before to after training at the group level. For example, researchers can ask whether training increases sensitivity to acoustic differences, strengthens category-level representations, or both, and whether these neural changes are accompanied by improved categorization performance. In a precision design, the focus shifts to how change unfolds within individuals. Researchers can ask when neural changes first emerge, whether sensitivity to acoustic features changes before category-level representations do, and how closely these neural changes track each individual's learning trajectory.

#### *5.1.6. Interpretation of results*

In a group-based design, results are summarized in group statistical brain maps, such as pre- versus post-training contrasts. Observed effects are interpreted as average training-related changes in neural representations across participants. Such findings support the inference that category learning is associated with neural change at the group level. In a precision design, results are instead visualized as brain maps and neural or behavioral trajectories for each participant (Miller et al., 2022; Newbold et al., 2020). Interpretation focuses on when and to what extent changes emerge within individuals. This shifts the interpretation to within-person trajectories and their replication across individuals.

### **5.2. Integrating group-based and precision approaches**

Individual-specific precision approaches are not optimal for every research question on learning-related plasticity. In some cases, the primary aim is to estimate effects at the population level rather than to characterize change within individuals, for example when evaluating whether a training intervention is effective on average in a clinical population. In other cases, dense longitudinal sampling may not be feasible because of practical, financial, or participant-related constraints. This may be especially relevant in studies involving clinical populations, children, or older adults, for whom the burden of repeated neuroimaging sessions may be too high.

In such settings, hybrid designs may provide a useful middle ground by combining advantages of group-based and precision approaches. For example, a study could acquire fMRI data at a limited number of time points, such as before, during, and after training, while still incorporating individualized functional localization of brain regions or networks of interest, analyses in native anatomical space, or more densely sampled behavioral measures. This approach preserves the possibility of estimating average effects at the population level while increasing sensitivity to individual functional neuroanatomy and brain-behavior relationships. However, compared with dense individual-specific designs, it is less well suited to capturing fine-grained within-person trajectories and nonlinear learning dynamics. Hybrid designs therefore offer a practical way to incorporate key elements of precision neuroimaging when fully dense sampling is neither necessary nor feasible. Ultimately, the most appropriate design depends on the inferential goal of the study and the level of temporal and individual detail required.

**Table 1. Comparison of group-based and precision designs for longitudinal studies of learning-related neural plasticity.**

The table summarizes idealized characteristics of the two design approaches. In practice, many studies combine elements of both, so the contrasts shown here are relative rather than absolute. Appropriate sample sizes depend on the target effect, namely group-average effects in group-based designs and individual-level effects in precision designs. Precision approaches typically prioritize reliable within-person estimation through dense sampling, whereas conventional group-based designs typically prioritize population-level estimation across larger samples.

| | ***Group-based design*** | ***Precision design*** |
|---|---|---|
| *Sample size* | Larger sample | Small sample |
| *Neuroimaging sessions* | Few | Many |
| *Baseline sampling* | Minimal | Repeated |
| *Stimulus sampling* | Narrower | Expanded |
| *Training period* | Short to moderate | Moderate to long |
| *Detectable temporal pattern* | Pre-post change | Linear and non-linear trajectories |
| *Primary question* | Whether change occurs | How change unfolds |
| *Analysis space* | Standard/common space | Native space |
| *Functional localization* | Group-defined | Individualized |
| *Brain-behavior analysis* | Across participants | Within participants |
| *Typical visualization* | Group contrast maps | Individual maps and trajectories |
| *Primary inference target* | Population-average effect | Within-person trajectories |
| *Replication unit* | Independent samples/studies | Individuals within study |

## 6. Conclusion

Individual-specific, multi-modal precision neuroimaging of learning-related plasticity offers a powerful framework to unmask core principles governing skill learning. Although this approach has only rarely been applied to the study of learning or to plasticity more generally, the emerging evidence (Chauvin et al., 2025; Miller et al., 2022; Newbold et al., 2020; Pritschet et al., 2020) highlights its considerable potential to generate novel insights and to advance cognitive neuroscience toward a more individualized understanding of brain function and learning.

## Glossary

**Aptitude-treatment interactions** describe the idea that the effects of training or an intervention depend on a person's initial characteristics, such as baseline ability. In learning research, this means that people with different starting points may show different amounts or forms of behavioral and neural change in response to the same training.

The **blood-oxygenation-level-dependent (BOLD) signal** is an indirect measure of brain activity used in functional MRI. It reflects changes in blood oxygen levels that occur when groups of neurons become more or less active.

**Degeneracy** refers to the fact that the same outcome or function can be achieved in different ways within the brain. This means that similar behavior or performance does not always rely on the exact same neural processes or brain regions.

**Denoising** refers to steps used to reduce unwanted noise in neuroimaging data, such as signals related to head motion or physiological processes, while preserving signals of interest. The goal of denoising is to improve data quality and make measurements more reliable.

**Extended sampling**, also called dense sampling, refers to collecting a large amount of data from the same individual, often across many trials, sessions, or timepoints.

**Hyperscanning** refers to measuring brain activity from two or more individuals at the same time, typically while they interact, communicate, or learn together.

**Measurement precision** describes how consistent a measurement is when it is repeated under the same conditions. High precision means that repeated measurements give very similar results, with little random variation.

**Multi-echo acquisition** refers to an fMRI method in which multiple images are acquired at different echo times following a single excitation. Comparing signals across echoes helps separate BOLD-related signal from non-BOLD noise, which can improve denoising and increase measurement reliability.

**Native space** refers to an individual's own anatomical brain space, in which neuroimaging data are analyzed without being transformed to a common template. It is often contrasted with standard space, which aligns data from different individuals to the same reference brain.

**Neural representations** refer to patterns of brain activity that are thought to carry information needed to perform a task, such as information about what we perceive or how we act. For example, in a longitudinal study of motor skill learning in humans, neural representations might be quantified as patterns of fMRI activity in motor cortex while participants perform finger movements, reflecting how specific finger movement sequences are represented in the brain.

**Non-ergodicity** refers to the concept that patterns observed across a group of individuals do not necessarily describe how a single individual behaves or changes over time.

**Population prevalence inference** refers to a statistical approach that estimates how common an effect is in a population based on how often it is observed across individuals in a study.

**Precision neuroimaging** refers to a family of approaches that use large amounts of neuroimaging data from each individual to precisely characterize brain function at the level of the individual. At present, there is no clear boundary between what counts as "precise" versus "imprecise".

**Skills** can be understood as abilities to carry out actions or judgments that improve with practice and experience. They can range from simple actions to more complex sequences, such as typing on a keyboard, recognizing animal sounds, playing a musical instrument, or solving a mathematical problem.

**Spatial normalization** is a common procedure in neuroimaging in which individual brain images are adjusted to match a standard brain template, such as the Montreal Neurological Institute (MNI) template. The aim is to enable comparisons of brain images across individuals using a shared coordinate system.

**Test-retest reliability** describes how similar measurements are when the same measure is taken again at a later time under comparable conditions. High test-retest reliability means that the measure remains stable across time, making it easier to tell whether observed changes reflect real changes over time.

## Acknowledgements

We thank Emily S. Cross, Richard Ramsey, and Luca A. Naudszus for providing valuable feedback on the manuscript. We thank Chantal Nagel for her expert assistance in creating the figure.

This work was supported by the Swiss National Science Foundation (Grants 222073 and 223775 to Simon Leipold) and the Social Brain Sciences Lab at ETH Zurich.

## Declaration of generative artificial intelligence (AI) use

During the preparation of this work, the corresponding author used ChatGPT (OpenAI) to improve the language and readability. The authors reviewed and edited the content as needed and take full responsibility for the content of the published article.